# Analysis of Thermal Stresses in Solidification of Spherical SLM Components

*Corresponding author:
Amir Mahyar Khorasani,
mahyar@deakin.edu.au

Academic editor:
Paul K. Collins

1-Amir Mahyar Khorasani, mahyar@deakin.edu.au,
2-Ian Gibson, ian.gibson@deakin.edu.au,
3-Moshe Goldberg, moshe.goldberg@deakin.edu.au,
4-Mohammad Masoud Movahedi
masoudmovahedi1969@yahoo.com
5-Guy Littlefair, guy.littlefair@deakin.edu.au
1-3 , 5 School of Engineering, Faculty of Science, Engineering and Built Environment, Deakin University, Geelong, Victoria, Australia
4- School of Science and Engineering, University of Applied Science and Technology, Qazvin, Iran

**Abstract**

Additive Manufacturing (AM) is the formalized term for what used to be called Rapid Prototyping and what is commonly referred to as 3D Printing. The key to how AM works is that parts are made by adding layers of material; each layer corresponding to a thin cross-section of the part derived from the original CAD data. Although most AM machines produce parts using polymers, there are an increasing number of machines that can directly fabricate in metals. The majority of these machines fabricate from raw material in powder form using a directed energy beam to create a local melt zone. Total hip replacement is recommended for people who have medical issues related to excessive wear of the acetabular, osteoarthritis, accident or age. Researches have shown that large numbers of hip arthroplasties (where the articular surface of a musculoskeletal joint is replaced), hip remodelling, or realignment are carried out annually and will increase in the next few decades. Manufacturing of acetabular shells by using AM is a promising and emerging method that has a great potential to improve public health. Lost wax casting or investment casting is currently used to produce acetabular shells followed by lengthy and complex secondary processes such as machining and polishing. Living organs and medical models have intricate 3D shapes that are challenging to identity in X-ray CT images. These images are used for preparing treatment plans to improve the quality of the surgeries regarding waiting and surgery time per procedure and care regime. For instance, a limited number of hip replacement procedures can be carried out on each acetabulum due to a decrease of bone thickness. Rapid prototyping is a suitable treatment planning tool in complex cases to enhance the quality of surgical procedure and provide long-term stability that can be used to customize the shape and size of the acetabular shell. In this paper, to analyse the manufacturing of a prosthetic acetabular shell, built-up lines resulting from a thermal stress flow and process stopping during the selective laser melting (SLM) AM process, with regarding Gibbs free energy, interfacial energy, and equilibrium temperature will be discussed. Geometrical measurements showed 1.59% and 0.27% differences between the designed and manufactured prototype for inside and outside diameter respectively. Experimental results showed that thermal stress flow in outer surfaces are compressive, but for inner surfaces are tensile, so built-up lines in inner and outer surfaces appear as a groove and dent respectively. The results also indicate that SLM is an accurate and promising method for fabrication of acetabular cup.

I. Introduction

Heat treatment in argon and vacuum environment with temperatures ranging from $750^0C$ to $1050^0C$ has been recommended to change microstructures by refining grain size or laminar structure and reduction of internal stress with optimized mechanical properties in metal parts. Mill annealing has been shown to decrease thermal shocks, stresses, and deformation which were common factors contributing to the failure. Furthermore, this heat treatment was shown to recrystallize the α phase and created an exclusive bi-modal microstructure consisting of coarse crab-claw-like primary α and fine lamellar transformed β phase which improved the quality of samples such as mechanical properties and uniformity (1-10). Investigation on microstructures of as-built Ti–6Al–7Nb that has been specified as a biomaterial according to ASTM standards showed that this alloy had α' martensite hardened by dispersive precipitates of the second phase when processed using SLM. The mentioned microstructure led to a reduction of ductility and increasing tensile and compressive strength compared to other fabrication processes. Also, heat treatment was recommended due to the high susceptibility of the as-built Ti—6Al—7Nb to prevent initiation of fatigue cracks (11).

Surface roughness plays an important role in the wear of an acetabular shell and its corresponding spherical femoral head. According to standards, the roughness (Ra) and roughness must lie within 0.02-0.036 µm and-and 0.9-7.3µm respectively to keep wear within acceptable limits. This is a critical factor in the material, construction, technology development and optimization of the acetabular shell (12, 13). Investigation on Co-Cr-Mo showed various factors such as clearance, head size, carbon content, and manufacturing method had an influence on metal-on-metal wear behaviour. This study proved that for an implant with a diameter of more than 36mm, if the lubrication was not complete for the whole surface, wear might increase. Smaller clearances and smoother surfaces may, therefore, decrease running-in wear (14-17).

Study on the manufacturing of human implant using AM showed considerable promise for the manufacturing of near net shape orthopedic implants with tailored material properties. Also, by increasing incident laser energy and decreasing scanning speed, high density (to near 100%) components was obtained. The Vickers hardness measurement on the acetabular shell also illustrated that hardness was highly related to the density as well as the production of (α') HCP martensite phase regimes within the process. Therefore, in AM of Ti-based alloys hardness and tensile strength increased ranging 37 to 57HRC and 0.9 to 1.45 Gpa respectively and break elongation decreased from 14% to 11% (18-22).

As a technology under development, the intricate SLM processes for producing Ti-6Al-4V parts are not yet fully understood. In this paper, thermal stress flow in the fabrication of acetabular shells that leads to visible lines by determining solidification on curvatures will be discussed.

   II. MATERIALS AND METHODS DESIGN AND MANUFACTURING ACETABULAR SHELL

The acetabular shell was designed using the SolidWorks parametric modelling software as a near net shape. This is illustrated in Figure 1 in different views. It has three holes to enable fixation onto the human bone and is part of the design that can be based on different people's skeleton to result in a better fit, kinematics, and comfort.

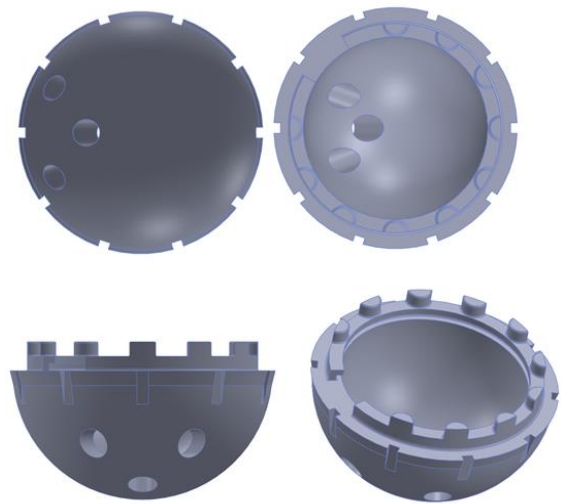

Figure 1 Designed acetabular shell using SolidWorks software

The equiaxed Ti-6Al-4V (Grade 5) powder was used as a stock material for the SLM machine that was produced by plasma-atomization process (Figure2). Samples were produced by using an SLM 125HL that is equipped with a YLR-Faser-Laser. Table 1 shows the system parameters.

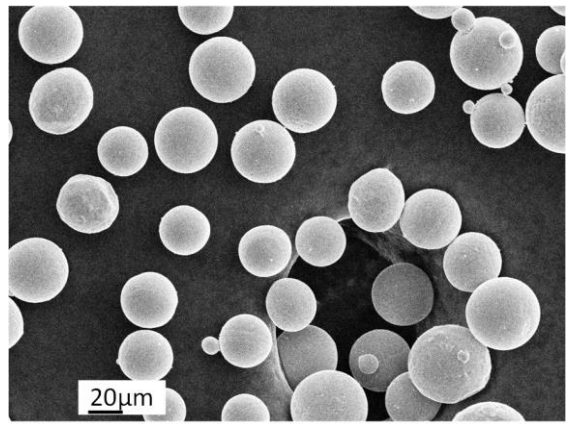

Figure 2 Ti-6Al-4V powdered for SLM

TABLE 1 SLM SYSTEM PARAMETERS

| System Parameters | Value |
|---|---|
| Laser Power | 100 W, YLR-Faser-Laser |
| Min. Scan Line / Wall Thickness | 120 µm |
| Operational Beam Focus Variable | 100 µm |
| Scan Speed | 700 mm/s |
| Hatch spacing | 75 µm |
| Layer thickness | 30 µm |
| Laser spot size | 0.2 mm |
| Inert Gas Consumption in Operation | Ar , 0,5 l/min |

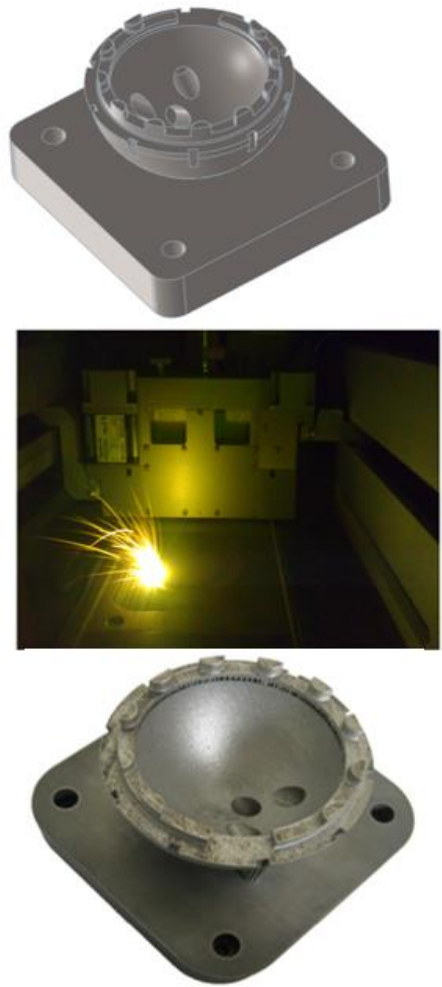

Figure 3 Fabrication process containing CAD and CAM.

Fabrication process containing CAD and cam process is shown in figure 3. The maximum dimensional deviation for printed sample is approximately 1.3μm and dimensional measurements showed 1.59% and 0.27% differences between the designed and produced samples for inside and outside diameter respectively. The measurements, therefore, indicated the accuracy of the SLM method for fabrication of acetabular shells to be acceptable for this application (13).

### III. THERMAL FLOW STRESS AND BUILT-UP LINE

Thermal flow stress and process stopping make visible distinguishing built-up lines on the surface of the sample that is solvable by machining or material removal processes (23). This problem can occur if there is a significant time interruption in the build process or for machines with smaller feeder hopper. Post processing to solve this problem leads to increase the size of printed sample and dimensional deviation. As can be seen in figure 5 the built-up line on the internal surface of the samples results in a groove line while a dent is observed on the outer surface. The reason for these phenomena in the fabrication of acetabular shell is related to the temperature in curvatures and thermal stress flow relations. To determine the process stopping the process and its effects on the surfaces of the acetabular shell, curvatures solidification theory such as planar and curved interfaces, kinematics of interfacial deformation, interfacial energy and Gibbs–Thomson effect are discussed. Based on experimental results mathematical relations can then be proved.

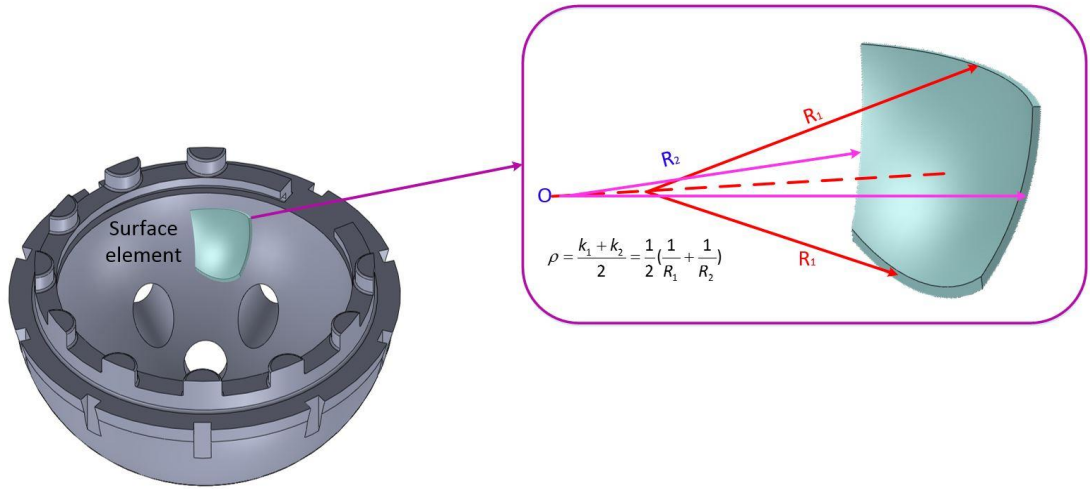

Figure 4 Mean curvature on surface element

Curvature, K, is a geometric characteristic of a plane curve which calculates the instant variations in the orientation angle Δθ of the normal vector with determining to the displacement along the curve's arc length, Δs that is $d\theta/ds$. In curvatures such as acetabular shell, two coefficients of the surface tensor that are defined along a surface coordinate are mean and Gaussian curvatures and are shown in equations 1-3 (24).

$$\begin{cases} [K_i] = \begin{bmatrix} k_1 & 0 \\ 0 & k_2 \end{bmatrix} = \begin{bmatrix} \frac{1}{R_1} & 0 \\ 0 & \frac{1}{R_2} \end{bmatrix} \Rightarrow \rho = \frac{k_1 + k_2}{2} = \frac{1}{2}(\frac{1}{R_1} + \frac{1}{R_2}) \\ \rho \equiv \frac{1}{2} tr[K_i] \end{cases} \quad (1)$$

Thickness plays an important role in solidification, thermal stress, and elongation in samples with the same inner radii. This phenomenon in acetabular shell surfaces is associated with the mean curvature, $\rho$, that relates to the change in the area, A, of a moving interface, to the volume, V, swept-out by a curved interface growing normal to itself (equation 2).

$$\rho = \frac{1}{2}(\frac{\partial A}{\partial V}) \quad (2)$$

The mean surface curvature, $\rho$, has a unit of (length)$^{-1}$, while Gaussian curvature that is defined according to equation 3 bears different unit (length)$^{-2}$. The geometrical coefficient, Ω, related to the rate of changing the spatial orientation (solid) angle, is defined by the following relation.

$$\kappa \equiv \det[k_i] = k_1 k_2$$
$$\kappa \equiv \frac{\partial \Omega}{\partial A} \quad (3)$$

Now by using governing relations on equilibrium at curved interfaces and Gibbs-Thomson effect, to analyse the value of thermal stress on the inner and outer surfaces of the acetabular shell, the equilibrium temperature is calculated. The Gibbs free energy is identified by, G, which is calculated from equation 4. In SLM process for fabrication of acetabular shell melting pressure in operation was constant so the volumetric and areal free energy coefficients were found on RHS of equation 4.

$$G(P,T) = U + PV - TS = H - TS \Rightarrow \frac{dG}{dt} =$$
$$\dot{G} = (\frac{\partial G_s}{\partial V_s})\frac{dV_s}{dt} + (\frac{\partial G_\ell}{\partial V_\ell})\frac{dV_\ell}{dt} + (\frac{\partial G_s}{\partial V_{s\ell}})\frac{dA_{s\ell}}{dt} \quad (4)$$

where U is the internal energy, P is pressure, V is volume, T is the temperature, S is the entropy and H are the enthalpy. Transformation from liquid to solid in constant pressure during acetabular shell fabrication leads to Gibbs free energy changes in each phase per unit volume, so equation five is derived:

$$\frac{\partial G_i}{\partial V_i} = -\frac{S_i}{\Omega_i}(T - T_{Ti_m}), (i = s, \ell) \tag{5}$$

where $T_{Ti_m}$ is Ti melting temperature, index of, s, is a solid phase and, $\ell$, is a liquid phase. The energy equal to the surface tension at an interface is called interfacial energy and is shown by $\gamma$ that is obtained by differentiating free energy of area (equation 6):

$$\frac{\partial G_i}{\partial A_s} = \gamma_{s\ell} \tag{6}$$

In freezing or melting of the acetabular shell, surface mass is conserved so the number of atoms moved to, or from, one phase $\dot{V}_s$ (solid/liquid) must be opposite, or the same to the number moved from, or to, the other (24). Equation 7 shows mass conservation for solidification in SLM process:

$$\frac{1}{\Omega_s}\frac{dV_s}{dt} = -\frac{1}{\Omega_\ell}\frac{dV_\ell}{dt} \tag{7}$$

The rate of $\dot{G}$ is associated with both area and volume changes that are naturally related to the deformation and motion of the curved surface illustrating the physical crystal-melt interface. If in this situation the variation of total free energy for a moving curved interface becomes zero then T is converted to equilibrium temperature ($T_{eq}$). By supposing monotone movement on the surfaces of acetabular shell $\dot{G}=0$ then by substitution equations 5 and six on Equation four the following relation is obtained (Equation 8):

$$\dot{G} = 0 = \left(-\frac{S_s}{\Omega_s}\right)(T_{eq} - T_{Ti_m})\frac{dV_s}{dt} + \left(-\frac{S_\ell}{\Omega_\ell}\right)(T_{eq} - T_{Ti_m})\frac{dV_\ell}{dt} + \gamma_{s\ell}\frac{dA_{s\ell}}{dt}$$
$$= \left[\left(-\frac{S_s}{\Omega_s}\right)(T_{eq} - T_{Ti_m}) + \left(-\frac{S_\ell}{\Omega_\ell}\right)(T_{eq} - T_{Ti_m}) + \gamma_{s\ell}\frac{dA_{s\ell}}{dV_s}\right]\dot{V}_s \tag{8}$$

In equation 8 the rate of melting or freezing has value, while free energy changing rate must be zero and in order to satisfy this requirement the amount of bracket must be zero, so by substituting equations 1 and 2 into equation 8 and determining $\Delta S_f = S_\ell - S_s$ as a molar entropy of fusion equation 9 is obtained for equilibrium temperature of acetabular shell fabrication on SLM process.

$$\frac{\Delta S_f}{\Omega_s}(T_{eq} - T_{Ti_m}) + \gamma_{s\ell}\frac{dA_{s\ell}}{dV_s} = 0 \Rightarrow$$
$$T_{eq} = T_{Ti_m} - \frac{2\gamma_{s\ell}\Omega_s}{\Delta S_s}\rho \Rightarrow T_{eq} = T_{Ti_m} - \frac{\gamma_{s\ell}\Omega_s}{\Delta S_s}\left(\frac{1}{R_1} + \frac{1}{R_2}\right) \tag{9}$$

Figures 1 and 3 show that the curvature surfaces of designed acetabular shell are symmetric so $R_1 = R_2$ then final equation for equilibrium temperature according to equation 10 is:

$$\begin{cases} T_{eq} = T_{Ti_m} - \frac{2\gamma_{s\ell}\Omega_s}{\Delta S_s R} \\ \frac{2\gamma_{s\ell}\Omega_s}{\Delta S_s R} = \xi \end{cases} \Rightarrow T_{eq} = T_{Ti_m} - \xi \tag{10}$$

From above equations, it is mentioned that by increasing radii (moving from inner surface to the outer) the value of $\xi$ decrease and RHS of equation 10 increases and subsequently, $T_{eq}$ increase. Therefore, according to equation 11 by determining $\Delta T = T_{eq} - T_r$ ($T_r$ is room temperature) thermal stress $\sigma_{th}$ increase by growing radii.

$$\sigma_{th} = \frac{E\alpha\Delta T}{1 - \upsilon} \tag{11}$$

where, E, is elastic modulus and $\upsilon$ is Poisson's coefficient. In other words, thermal stress on curvatures flows from inner surfaces to the outer surfaces. This phenomenon is proved by determining built-up lines on Figure 5.

Tensile stresses lead to groove on the inner surfaces, while compressive pressure on the outer surfaces transfer extra powder to the outer surface and small dent with 100μm is observed in Figure 5 which is 0.15% of the printed sample with 70mm diameter.

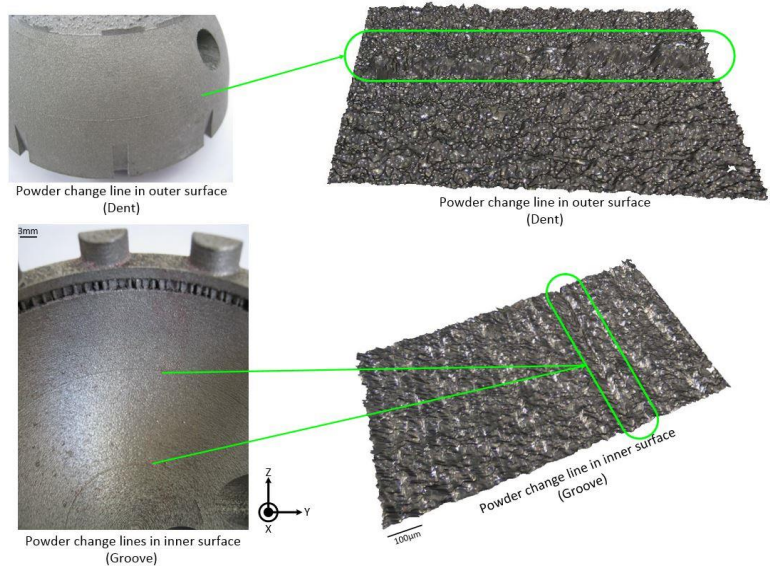

Figure 5 Process stopping line

### IV. RESULTS AND DISCUSSION
**BUILT UP LINE**

According to equation 4 Gibbs free energy is related to entropy and by determining Q, as transferred heat into the system, $C_P$ as specific heat capacity and m as mass from equation 12 entropy are calculated:

$$\begin{cases} S = \int_0^{T_{Ti_m}} C_p dT \\ C_p = Q_i / m\Delta T \Rightarrow S = \frac{Q_i}{m} LnT \\ i = s, \ell \end{cases} \quad (12)$$

By substitution equation 12 on 4 and six equation 13 is obtained:

$$\gamma_{s\ell} = \frac{\partial(u + PV - T\frac{QLnT}{m})}{\partial A_s} \quad (13)$$

Now by increasing mass (outer surfaces) through small changes for density Gibbs free energy and subsequently interfacial energy increase. Interfacial energy quantifies the disruption of intermolecular bonds that occur when a surface is created so by increasing interfacial energy according to equation 10, $\xi$ increases. Molar entropy of fusion is obtained from equation 14 (25, 26):

$$\Delta S_f = S_\ell - S_s = \frac{\Delta H}{T_{Ti_m}} + \int_{T_{Ti_m}}^{T_v} C_p(\ell) dT - \int_{T_r}^{T_{Ti_m}} C_p(s) dT \quad (14)$$

where, $T_v$ is vaporization temperature and $C_p$ is specific heat capacity. By substituting equation 12 on 14 molar fusion entropy is defined according to the following relation:

$$\Delta S_f = \frac{\Delta H}{T_{Ti_m}} + \frac{1}{m}\left( Q_\ell Ln \frac{T_v}{T_{Ti_m}} - Q_s Ln \frac{T_{Ti_m}}{T_r} \right) \quad (15)$$

equation 16 is derived from equation 10, 11 and 13-15 and shows by increasing mass (outer surface) as mentioned $\xi$ increases. By determining equation 16, the growth rate (the power) of m, in numerator and denominator is the same but the value of radii R, increases in the outer surface, therefore, $\xi$ is decreased, and equilibrium temperature and thermal stresses increased. Consequently, it is understood that due to thermal stresses and interfacial energy the flow of stress moves from the inner surface (tensile) to outer (compressive) so visible lines appear on the surface of produced prototypes.

$$\frac{2\Omega_s}{\frac{\Delta H}{T_{Ti_m}} + \frac{1}{m}\left(Q_\ell Ln\frac{T_v}{T_{Ti_m}} - Q_s Ln\frac{T_{Ti_m}}{T_r}\right)R} \frac{\partial(u + PV - T\frac{QLnT}{m})}{\partial A_s} = \xi \quad (16)$$

$$\sigma_{th} = \frac{E\alpha}{1-\upsilon}(T_{Ti_m} - \xi - T_r)$$

## V. CONCLUSION

SLM has high compatibility with 3D drawing, production of near net shape, flexibility in shape, ability to include porosity in the design, good hardness properties, geometrical complexity, and ability to customise to individual patient data. These pros characterize it as one of the promising processes in the production of prosthetic organs such as acetabular shells.

In the SLM process, due to a high melting temperature of Ti-6Al-4V, large thermal gradients occur which lead to the build-up of thermal shocks, elongation, and stress, while solidification flow is a segregated phenomenon that increases non-equilibrium phases. Thermal stress flow leads to visible lines on inner and outer surfaces. Curvature surfaces on acetabular structure have especial behaviours in solidification, and experimental results showed that thermal stress flow in outer surfaces is compressive, while in inner surfaces are tensile which made visible lines and grooves with about 100μm. To remove these lines post processing such as machining is needed that can increase the risk of dimensional deviation, therefore, it is recommended that the original size of the sample is designed 0.15% smaller.